\documentclass[aps,prl,twocolumn,superscriptaddress,longbibliography,ctexart]{revtex4-1}
\usepackage{graphicx}
\usepackage{latexsym}
\usepackage{amssymb}
\usepackage{amsmath}
\usepackage{amsfonts}
\usepackage{upgreek}
\usepackage{float}
\usepackage{bm}
\usepackage{multirow}
\usepackage{color}
\usepackage[T1]{fontenc}
\usepackage{hyperref}
\usepackage{subfigure}
\usepackage{booktabs,tabularx}
\usepackage{CJK}
\usepackage{diagbox}
\usepackage{array}

\hypersetup{
colorlinks = true,
linkcolor = [rgb]{0.70,0.13,0.13},
citecolor = [rgb]{0.13,0.55,0.13},
urlcolor  = [rgb]{0.25, 0.41, 0.88}}

\begin{document}
\title{Fidelity and criticality in the nonreciprocal Aubry-Andr{\'e}-Harper model}
\author{Chen-Chang Zeng}
\affiliation{College of Physics, Nanjing University of Aeronautics and Astronautics, Nanjing, 211106, China}

\author{Zhen Cai}
\affiliation{College of Physics, Nanjing University of Aeronautics and Astronautics, Nanjing, 211106, China}

\author{Guang-Heng Wang}
\affiliation{College of Physics, Nanjing University of Aeronautics and Astronautics, Nanjing, 211106, China}

\author{Gaoyong Sun}
\thanks{Corresponding author: gysun@nuaa.edu.cn}
\affiliation{College of Physics, Nanjing University of Aeronautics and Astronautics, Nanjing, 211106, China}
\affiliation{Key Laboratory of Aerospace Information Materials and Physics (NUAA), MIIT, Nanjing 211106, China}

\begin{abstract}
We study the critical behaviors of the ground and first excited states in the one-dimensional nonreciprocal Aubry-Andr{\'e}-Harper model using both the self-normal and biorthogonal fidelity susceptibilities. We demonstrate that fidelity susceptibility serves as a probe for the phase transition in the nonreciprocal AAH model. For ground states, characterized by real eigenenergies across the entire regime, both fidelity susceptibilities near the critical points scale as $N^{2}$, akin to the Hermitian AAH model. However, for the first-excited states, the fidelity susceptibilities exhibit distinct scaling laws, contingent upon whether the lattice consists of even or odd sites. For even lattices, both the self-normal and biorthogonal fidelity susceptibilities near the critical points continue to scale as $N^{2}$. In contrast, for odd lattices, the biorthogonal fidelity susceptibilities diverge, while the self-normal fidelity susceptibilities exhibit linear behavior, indicating a novel scaling law.

\end{abstract}

\maketitle

{\it Introduction.-} Non-Hermitian systems have become a prominent focus of physics research in recent years, thanks to their distinctive physical properties without the counterparts in Hermitian systems \cite{bergholtz2021exceptional, ashida2021non}.
The well-known fascinating phenomena are non-Hermitian skin effects \cite{lee2016anomalous,yao2018edge,kunst2018biorthogonal,gong2018topological,yokomizo2019non,yang2020non,okuma2020topological,zhang2020correspondence,wang2020defective,jiang2020topological,weidemann2020topological,xiao2020non,alvarez2018non,borgnia2020non,lin2023topological,geng2023nonreciprocal,shao2024non}, 
exceptional points \cite{heiss2012physics,hodaei2017enhanced,miri2019exceptional,yang2019non,yoshida2019symmetry,zhou2018observation,dora2019kibble,jin2020hybrid,xiao2021observation,chen2023switchable,zhang2023controlling,li2023maximum},
and continuous quantum phase transition without gap closing \cite{matsumoto2020continuous,yang2022hidden}.
Recent findings have demonstrated that the eigenstates of non-Hermitian systems exhibit distinct behaviors \cite{zhang2022symmetry,zhai2022nonequilibrium,lu2024many,lu2024manyIsing},
highlighting the crucial importance of understanding phases and phase transitions within this realm.

The Aubry-Andr{\'e}-Harper (AAH) model is an intriguing simple model employed for studying the phase transition from the extended to localized phases \cite{aubry1980analyticity,harper1955single,kraus2012topological} or topological phases \cite{lang2012edge} in Hermitian systems.
Recently, the AAH model has been extended to incorporate non-Hermitian systems \cite{zhai2022nonequilibrium,jiang2019interplay,zeng2020topological,wu2021non,longhi2021phase,liu2021exact,cai2021boundary,zhai2021cascade,liu2021localization,zeng2022real,tang2022topological,dai2022dynamical,chen2022quantum,han2022dimerization,lin2022topological,dai2023emergence,padhi2024quasiperiodic}, 
providing a platform for exploring the interplay between non-Hermiticity and quasiperiodicity, where the phase transition between delocalization and localization \cite{zhai2022nonequilibrium,jiang2019interplay} as well as the topological phases \cite{zeng2020topological} persist. 
Moreover, a $\mathcal{PT}$ transition is expected to take place within the non-Hermitian AAH model \cite{zhai2022nonequilibrium,jiang2019interplay}.
Nevertheless, it has been observed that certain eigenstates, such as the ground state, can remain real \cite{zhang2022symmetry,zhai2022nonequilibrium,lu2024many,lu2024manyIsing} even as they cross the critical points (exceptional points) of the $\mathcal{PT}$ transitions. 
A fundamental question arises: Can the eigenstate, characterized by the real eigenenergies across the entire regime of a parameter, adequately describe the $\mathcal{PT}$ transition?

\begin{figure}[t]
	\centering
	\includegraphics[width=9.5cm]{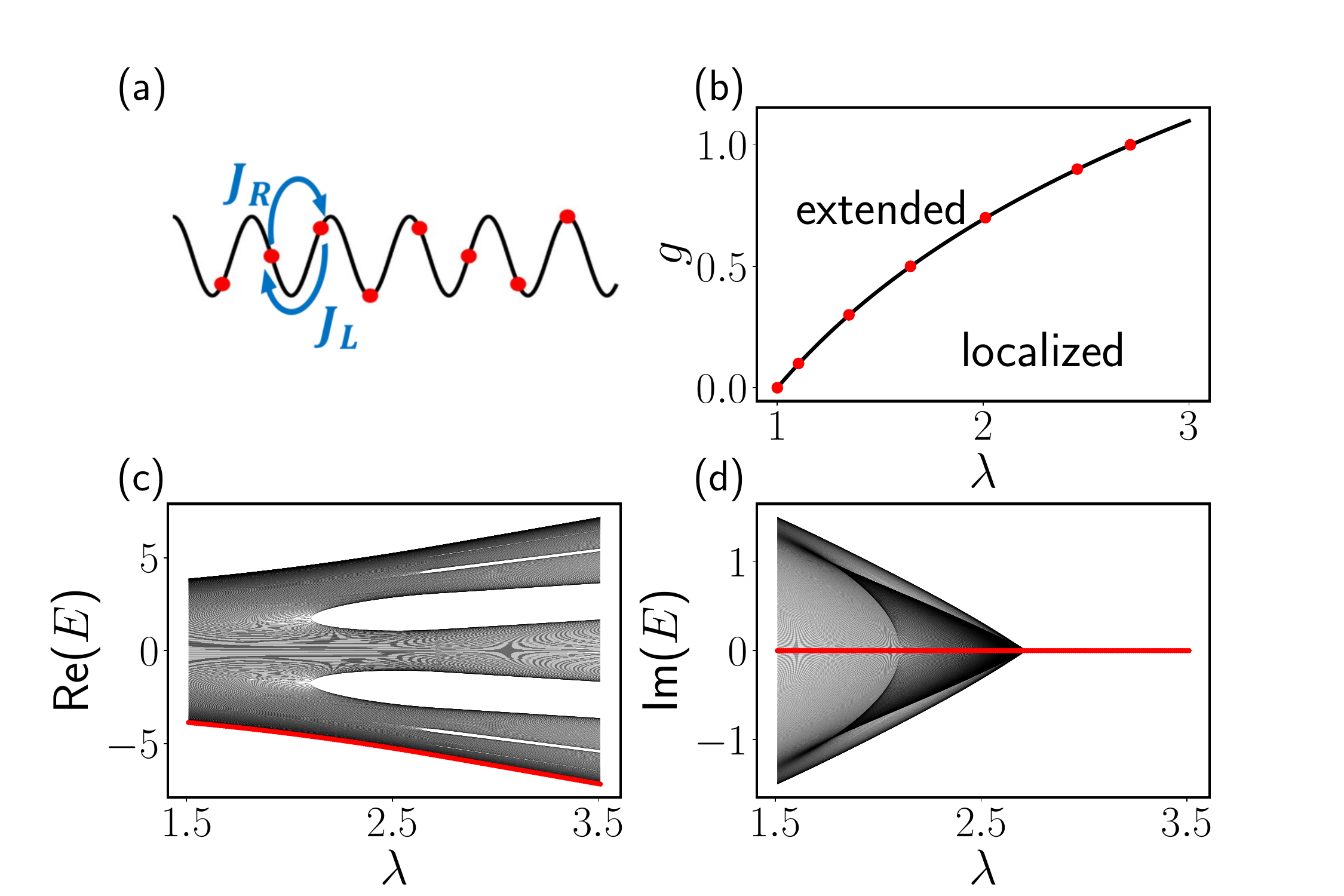}
	\caption{(a) Sketch of the one-dimensional nonreciprocal AAH model with the numbers of the lattice $N=8$. 
	(b) The phase transition in PBCs with respect to $g$ and $\lambda$, the circle symbols and the solid line denote the numerical results of the critical points and analytical critical line $\lambda = e^g$, respectively.
	(c) The real part of energies with respect to $\lambda$ for $N=987$ at $g=1$.
	(d) The imaginary part of energies with respect to $\lambda$ for $N=987$ at $g=1$. The red solid lines represent the ground state of the AAH model.}
	\label{fig:PhaseDiagram}
\end{figure}

The exploration of phase transitions in non-Hermitian systems has emerged as a significant focus in recent years within non-Hermitian research. Consequently, characterizing these transitions is critically important within this domain.
A simple method for studying quantum phase transitions in Hermitian systems involves analyzing the fidelity susceptibility \cite{Zanardi2006,You2007,Venuti2007,Chen2008,Gu2008,Yang2008,Gu2010,Albuquerque2010,Sun2017,Zhu2018,Luo2018}, 
facilitated by the advancements in quantum information science.
Fidelity susceptibility, a proven valuable tool for studying quantum phase transitions in Hermitian systems, has recently been extended to non-Hermitian systems \cite{matsumoto2020continuous,jiang2018topological,zhang2019quantum,tzeng2021hunting,sun2022biorthogonal,tu2023general}. 
It has been observed that the biorthogonal fidelity susceptibilities can describe the phase transition of a non-Hermitian system in the real-energy regime with the same scaling law as in Hermitian systems \cite{sun2022biorthogonal}, while the fidelity instead approaches a constant value of 1/2 near the exceptional points \cite{tu2023general}. In contrast, the scaling laws of the self-normal fidelity susceptibility have been less thoroughly studied \cite{matsumoto2020continuous}.
For the specific Hermitian AAH model, fidelity susceptibility has been employed to characterize transitions from delocalization to localization \cite{wei2019fidelity,cookmeyer2020critical,lv2022quantum,lv2022exploring}, wherein the system is argued to segregate into three subsequences \cite{cookmeyer2020critical,lv2022exploring} contingent upon the odd/even lattice structure.
However, the use of fidelity to study the non-Hermitian AAH model remains unexplored.
An intriguing question arises: Can fidelity susceptibility effectively characterize phase transitions of the non-Hermitian AAH model? Moreover, should fidelity susceptibility be segmented into three subsequences?

Motivated by these questions, we study the quantum criticality, particularly at the exceptional points, of non-Hermitian AAH model 
using both the self-normal and biorthogonal fidelity susceptibilities of the ground state and the first-excited state.
We find that the fidelity susceptibility exhibits smooth changes when the energies are real, yet undergoes significant alterations when the energy of the system transitions from real to complex.
For ground states, it is found that both fidelity susceptibilities near the critical points scale as $N^{2}$, indicating a second-order phase transition. In contrast, for first-excited states, the fidelity susceptibilities exhibit distinct scaling laws. The self-normal fidelity susceptibilities scale as $N^{2}$ for even lattices, but exhibit linear behaviors for odd lattices where $\mathcal{PT}$ transitions occur, revealing a new scaling law that has not been previously discussed \cite{matsumoto2020continuous}.


\begin{figure}[t]
	\centering
	\includegraphics[width=8.9cm]{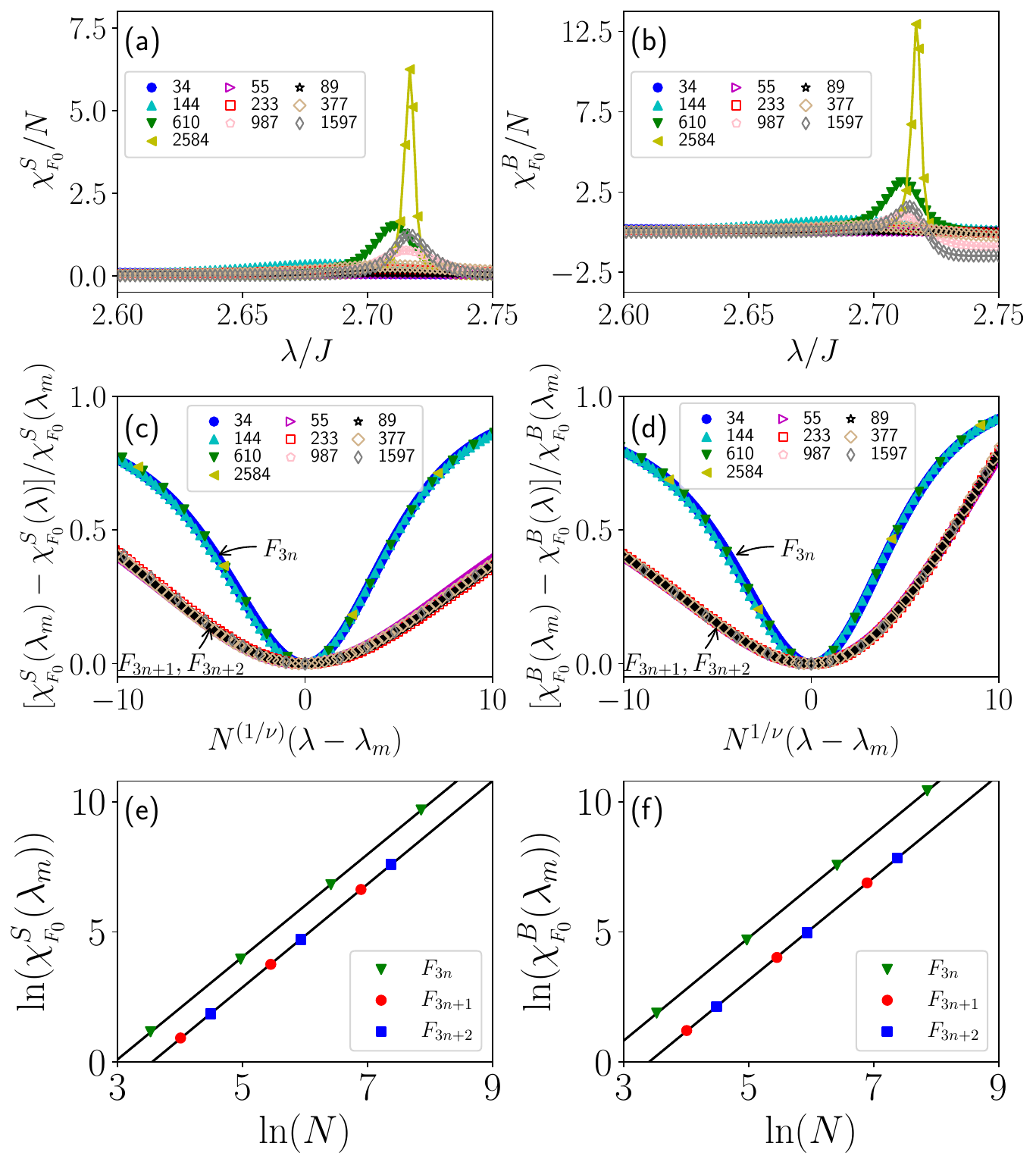}
	\caption{ Finite-size scalings of ground-state fidelity susceptibilities.
	(a)(b) The self-normal fidelity susceptibility per site $\chi^S_{_{F_0}}/N$ and the biorthogonal fidelity susceptibility per site $\chi^B_{_{F_0}}/N$ as functions of $\lambda$ for $N=34, 55, 89, 144, 233, 377, 610, 987,1597, 2584$.  
	(c)(d) Data collapses of $\left[\chi_{F_{0}}^{S, B}\left(\lambda_m\right)-\chi_{F_{0}}^{S, B}(\lambda)\right] / \chi_{F_{0}}^{S, B}\left(\lambda_m\right)$ 
	as a function of $N^{(1 / \nu)}\left(\lambda-\lambda_m\right)$, 
	where $\nu=1.014$ and $\nu=1.016$ for $\chi^S_{_{F_0}}$ are obtained from odd and even lattices;  $\nu=0.995$ and $\nu=1.011$ for $\chi^B_{_{F_0}}$ are obtained from odd and even lattices.
	(e)(f) The finite-size scaling behaviors of $\ln(\chi^S_{_{F_0}}(\lambda_m))$ and $\ln(\chi^B_{_{F_0}}(\lambda_m))$,
	where $\nu=1.008 \pm 0.0056$ and $\nu=1.015 \pm 0.0095$ for $\chi^S_{_{F_0}}$ are derived from odd and even lattices;  $\nu=1.014 \pm 0.0071$ and $\nu=1.012 \pm 0.0068$ for $\chi^B_{_{F_0}}$ are derived from odd and even lattices.
	}
	\label{fig:E0all}
\end{figure}

{\it Model.-} In the following, we consider the one-dimensional nonreciprocal AAH model as shown in Fig.\ref{fig:PhaseDiagram}(a).
The Hamiltonian is given by \cite{zhai2022nonequilibrium,jiang2019interplay, zhai2021cascade},
\begin{align}
H=\sum_{j=1}^{N} -\left(J_{L} c_{j}^{\dagger} c_{j+1}+J_{R} c_{j+1}^{\dagger} c_{j}\right)+2 \lambda \cos (2 \pi \alpha j)c_{j}^{\dagger} c_{j},
\label{eq:HamAAH}
\end{align}
where $c_{j}^{\dagger}$ and $c_{j}$ represent the fermionic creation and annihilation operators at the $j$th lattice site respectively. 
Here, $N$ denotes the length of the chain;
$J_{L}=J e^{-g}$ and $J_{R}=J e^{g}$ are the left and right hopping integrals between two adjacent lattices, 
which are nonreciprocal hopping terms introducing a non-Hermiticity;
$\lambda$ is the amplitude of the chemical potential. The quasiperiodicity is imposed by the irrational number $\alpha$, referred to as the golden rate $(\sqrt5-1)/2$. 
In a finite lattice, the ratio of fibonacci series $F_{n-1}/F_n$ can substitute for the golden rate $\alpha$ when the lattice size is $N = F_n$. 

When $g=0$, the system in Eq.(\ref{eq:HamAAH}) is the conventional Hermitian AAH model, which undergoes a phase transition from the extended phase to the localized phase at $\lambda=1$ \cite{wei2019fidelity} [c.f. Fig.\ref{fig:PhaseDiagram}(b)].
Interestingly, the system can be divided into three subsequences, $F_{3n} = 8,34,144,610,2584,...$, $F_{3n+1} = 13,55,233,987,...$, and $F_{3n+2} = 21,89,377,1597,...$,
dependent on the odd/even properties of the denominators $F_{n}$ in the irrational number $\alpha$ \cite{cookmeyer2020critical}. 
The subsequence $F_{3n}$ is even, while the other two subsequences are odd. 
When $g \neq 0$, the system in Eq.(\ref{eq:HamAAH}) represents the non-Hermitian AAH model due to the nonreciprocal hopping, 
undergoing a phase transition from the extended phase to the localized phase at $\lambda=e^{g}$ \cite{zhai2022nonequilibrium,jiang2019interplay, zhai2021cascade}. 
Throughout the paper, we investigate the characteristics of the nonreciprocal AAH model by employing fidelity susceptibility with $J=1$ under periodic boundary conditions (PBCs).


\begin{figure}[t]
	\centering
	\includegraphics[width=8.9cm]{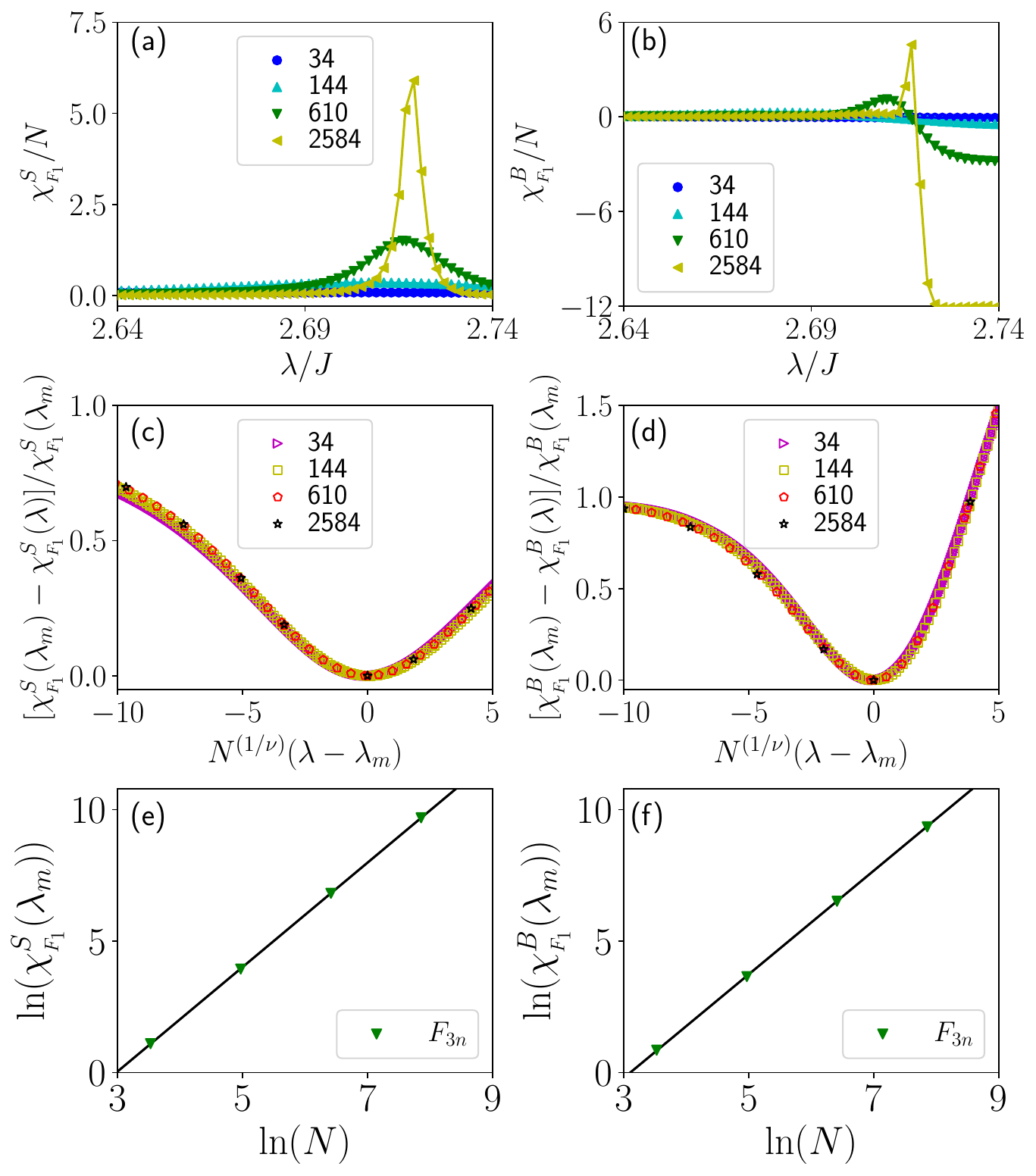}
	\caption{Finite-size scalings of the first-excited state fidelity susceptibilities for even lattices.
	(a)(b) The self-normal fidelity susceptibility $\chi^S_{_{F_1}}/N$ and the biorthogonal fidelity susceptibility $\chi^B_{_{F_1}}/N$ as functions of $\lambda$ for $N=34, 144, 610, 2584$.  
	(c)(d) Data collapses of $\left[\chi_{F_{1}}^{S, B}\left(\lambda_m\right)-\chi_{F_{1}}^{S, B}(\lambda)\right] / \chi_{F_{1}}^{S, B}\left(\lambda_m\right)$ 
	as functions of $N^{(1 / \nu)}\left(\lambda-\lambda_m\right)$, 
	where $\nu=1.015$ for $\chi^S_{_{F_1}}$ and $\nu=0.996$ for $\chi^B_{_{F_1}}$.
	(e)(f) The finite-size scaling behaviors of $\ln(\chi^S_{_{F_1}}(\lambda_m))$ and $\ln(\chi^B_{_{F_1}}(\lambda_m))$,
	where $\nu=1.007 \pm 0.0047$ and $\nu=1.015 \pm 0.0058$ are obtained for $\chi^S_{_{F_1}}$ and $\chi^B_{_{F_1}}$, respectively.
	}
	\label{fig:E1even}
\end{figure}

{\it Fidelity susceptibility.-} In a quantum system, two states can exhibit a similarity within the same phase, even when the interval of the controlled parameter between them is large.
Around the critical point of the phase transition, despite a small parameter gap between the two states, they can exhibit an increased overlap enhanced by the quantum criticality \cite{quan2006decay}.
For a non-Hermitian system, quantum fidelities are characterized by two definitions: the self-normal fidelity and the biorthogonal fidelity, depending on the usage of the eigenstates \cite{sun2022biorthogonal}.

The self-normal fidelity is defined as the overlap between two right eigenstates corresponding to distinct parameters,
\begin{equation}
F^{S}_{n}(\lambda, \delta \lambda) =\frac{\left\langle\psi_n^{R}(\lambda) \mid \psi_n^{R}(\lambda+\delta \lambda)\right\rangle}
{\left\langle\psi_n^{R}(\lambda) \mid \psi_n^{R}(\lambda)\right\rangle \left\langle\psi_n^{R}(\lambda+\delta \lambda) \mid \psi_n^{R}(\lambda+\delta \lambda)\right\rangle},
\label{eq:F}
\end{equation}
where $|\psi_n^{R}(\lambda) \rangle$ is the $n$th right eigenstate of the AAH model at $\lambda$.
The biorthogonal fidelity is defined as the inner product of both the left and the right eigenstates corresponding to two distinct parameters \cite{tzeng2021hunting,sun2022biorthogonal},
\begin{equation}
F^{B}_{n}(\lambda, \delta \lambda) =\frac{\left\langle\psi_n^L(\lambda) \mid \psi_n^R(\lambda+\delta \lambda)\right\rangle \left\langle\psi_n^L(\lambda+\delta \lambda) \mid \psi_n^R(\lambda)\right\rangle}
{\left\langle\psi_n^L(\lambda) \mid \psi_n^R(\lambda)\right\rangle \left\langle\psi_n^L(\lambda+\delta \lambda) \mid \psi_n^R(\lambda+\delta \lambda)\right\rangle},
\label{eq:FLR}
\end{equation}
where $|\psi_n^{L}(\lambda) \rangle$ is the $n$th left eigenstate of the AAH model at $\lambda$.
The values of the fidelities $F^{S}_{n}(\lambda, \delta \lambda)$ and $F^{B}_{n}(\lambda, \delta \lambda)$ dependent on the controlled parameters $\lambda$ and $\delta\lambda$. 
In the Taylor expansion of the fidelity, the main contribution originates from the second derivative term, and the coefficient associated with this term is termed the fidelity susceptibility, defined as:
\begin{equation}
\chi_{F}(\lambda)=\lim _{\delta \lambda \rightarrow 0} \frac{-2 \ln |F(\lambda, \delta \lambda)|}{(\delta \lambda)^2} = \lim _{\delta \lambda \rightarrow 0} \frac{2(1-F(\lambda, \delta \lambda))}{(\delta \lambda)^2}
\label{eq4}
\end{equation}
For second-order transitions, the fidelity susceptibility scales as \cite{wei2019fidelity},
\begin{equation}
\chi_{F}(\lambda) = N^{2/\nu} \Phi((\lambda -\lambda_m)N^{1/\nu}),
\label{eq:FScollpase}
\end{equation}
from the finite-size scaling theory, where $\nu$ is the correlation length critical exponent and $\lambda_m$ denotes the peak position of the fidelity susceptibility.
Moreover, the fidelity susceptibility also exhibits a scaling law \cite{Gu2010,Albuquerque2010,Sun2017,Zhu2018},
\begin{equation}
\chi_{F}(\lambda_{m}) \propto N^{2/\nu},
\label{eq:FSscaling}
\end{equation}
near the critical point.
Thus one can simply determine the correlation length critical exponent $\nu$ from Eq.(\ref{eq:FSscaling}).
However, the scaling laws of fidelity susceptibility are less known for the $\mathcal{PT}$ transitions.

In the following, we primarily employ the self-normal fidelity susceptibility $\chi^S_{F_n}$ and the biorthogonal fidelity susceptibility $\chi^B_{F_n}$ of the ground state and the first-excited state 
to investigate phase transitions, especially the $\mathcal{PT}$ transitions, of the one-dimensional nonreciprocal AAH model as described in Eq.(\ref{eq:HamAAH}).

\begin{figure}[t]
	\centering
	\includegraphics[width=8.9cm]{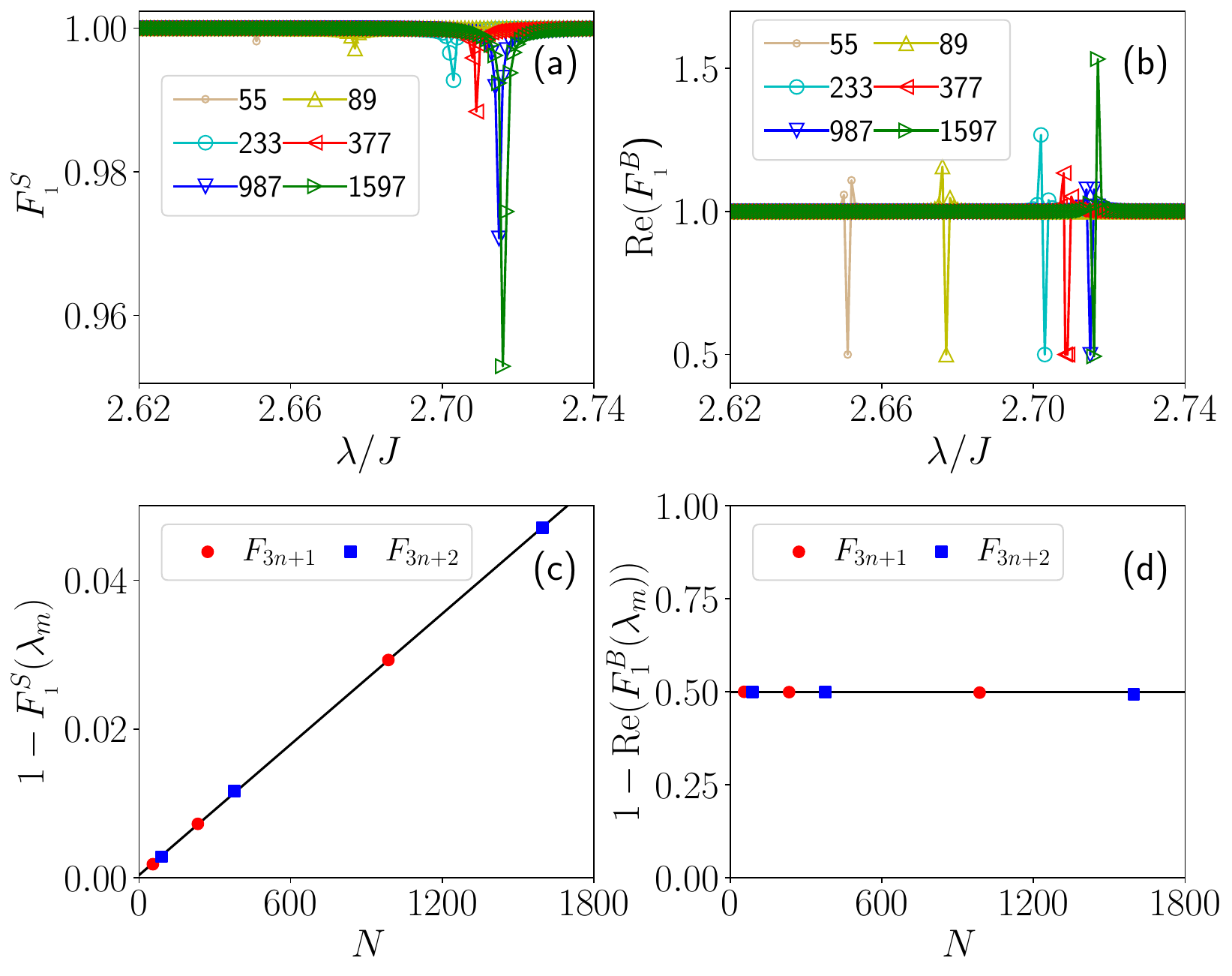}
	\caption{ First-excited state fidelities for odd lattices.
	(a) (b) The self-normal fidelity $F_1^{S}$ and the real-part values of the biorthogonal fidelity $\text{Re}(F_1^{B})$ as a function of $\lambda$ for $N=55,89,233,377,987,1597$.  
	(c) (d) The scaling behaviors of the self-normal fidelity $1-F_1^{S}(\lambda_m)$ and the real part of the biorthogonal fidelity $1- \text{Re}(F_1^{B}(\lambda_m))$, with the slope of $0.9721 \pm 0.0050$ and the slope of $-0.0036 \pm 0.0031$, respectively. Note that the fidelity scales as $\chi_{F}(\lambda_{m}) \propto (1-F(\lambda_{m})) / (\delta \lambda)^2$ for small $\delta \lambda$ according to Eq.(\ref{eq4}).
	}
	\label{fig:E1odd}
\end{figure}

{\it Phase transitions.-} We employ exact diagonalization to explore the nonreciprocal AAH model, unveiling the phase diagram depicted in Fig.\ref{fig:PhaseDiagram}(b).
The phase diagram is derived from the real-complex transition [c.f. Fig.\ref{fig:PhaseDiagram}(c)(d)], thereby characterizing it as a $\mathcal{PT}$ transition.
As illustrated in Fig.\ref{fig:PhaseDiagram}, the system demonstrates non-zero imaginary components within the energy spectrum around $\lambda_c \approx 2.71$, 
slightly deviating from the analytical value $\lambda_c = e^g$ at $g=1$  \cite{zhai2022nonequilibrium,jiang2019interplay, zhai2021cascade} due to the finite-size effect.
As the system size increases, the critical point tends toward $\lambda_c = e$ \cite{zhai2022nonequilibrium,jiang2019interplay, zhai2021cascade}. 

\begin{table}[!t]
    \centering
    \renewcommand{\arraystretch}{1.5}
    \caption{Scaling laws of fidelity susceptibility in the nonreciprocal AAH model near the critical points.}
	
    \begin{tabular}{|c|c|c|c|c|}
   \hline 
    \textbf{States} &  \multicolumn{2}{c|}{Ground states}  &  \multicolumn{2}{c|}{1st excited states} \\ \hline
   \textbf{Fidelity susceptibility} & ~~$\chi^B_{_{F_0}}$ ~~~& $\chi^S_{_{F_0}}$ & ~~~$\chi^B_{_{F_1}}$ ~~~& $\chi^S_{_{F_1}}$ \\ \hline
   \textbf{Even lattice}  & $N^2$   & $N^2$   & $N^2$   & $N^2$ \\ \hline
   \textbf{Odd lattice}   & $N^2$   & $N^2$    & $-\infty$  & $N$ \\ 
   \hline
   \end{tabular}
   \label{tab:scalingFS}
\end{table}

Interestingly, it was found that the ground-state energy of the AAH model changes smoothly and maintains its real value despite the $\mathcal{PT}$ transition occurring  \cite{zhai2022nonequilibrium,jiang2019interplay, zhai2021cascade}.
To examine whether the ground-state fidelity susceptibilities can accurately depict the transition,
we calculate both the self-normal fidelity susceptibility $\chi_{F_0}^{S}$ and the biorthogonal fidelity susceptibility and $\chi_{F_0}^{B}$ of the ground state, as depicted in Fig.\ref{fig:E0all}.
As can be seen from Fig.\ref{fig:E0all}, both the self-normal fidelity susceptibility and the biorthogonal fidelity susceptibility per site increase [c.f. Fig.\ref{fig:E0all}(a)(b)], 
indicating a phase transition in the ground state.
We observe that both the self-normal fidelity susceptibility and the biorthogonal fidelity susceptibility collapse into two distinct curves [c.f. Fig.\ref{fig:E0all}(c)(d)] depending on the odd/even lattice.
The correlation length critical exponent $\nu \approx 1$ is determined, consistent with the Hermitian AAH model \cite{zhai2022nonequilibrium}.
Additionally, the correlation length critical exponent ($\nu = 1$) is doubly verified by fitting the maximum values of both $\chi_{F_0}^{S}$ and $\chi_{F_0}^{B}$ [c.f. Fig.\ref{fig:E0all}(d)(e)].
Our findings suggest that the ground-state fidelity susceptibility serves as an excellent tool for detecting the phase transition of the nonreciprocal AAH model. 
Moveover, we show that the ground-state phase transition of the nonreciprocal AAH model is a second-order phase transition rather than a $\mathcal{PT}$ transition \cite{zhai2022nonequilibrium}.

\begin{figure}[t]
	\centering
	\includegraphics[width=8.6cm]{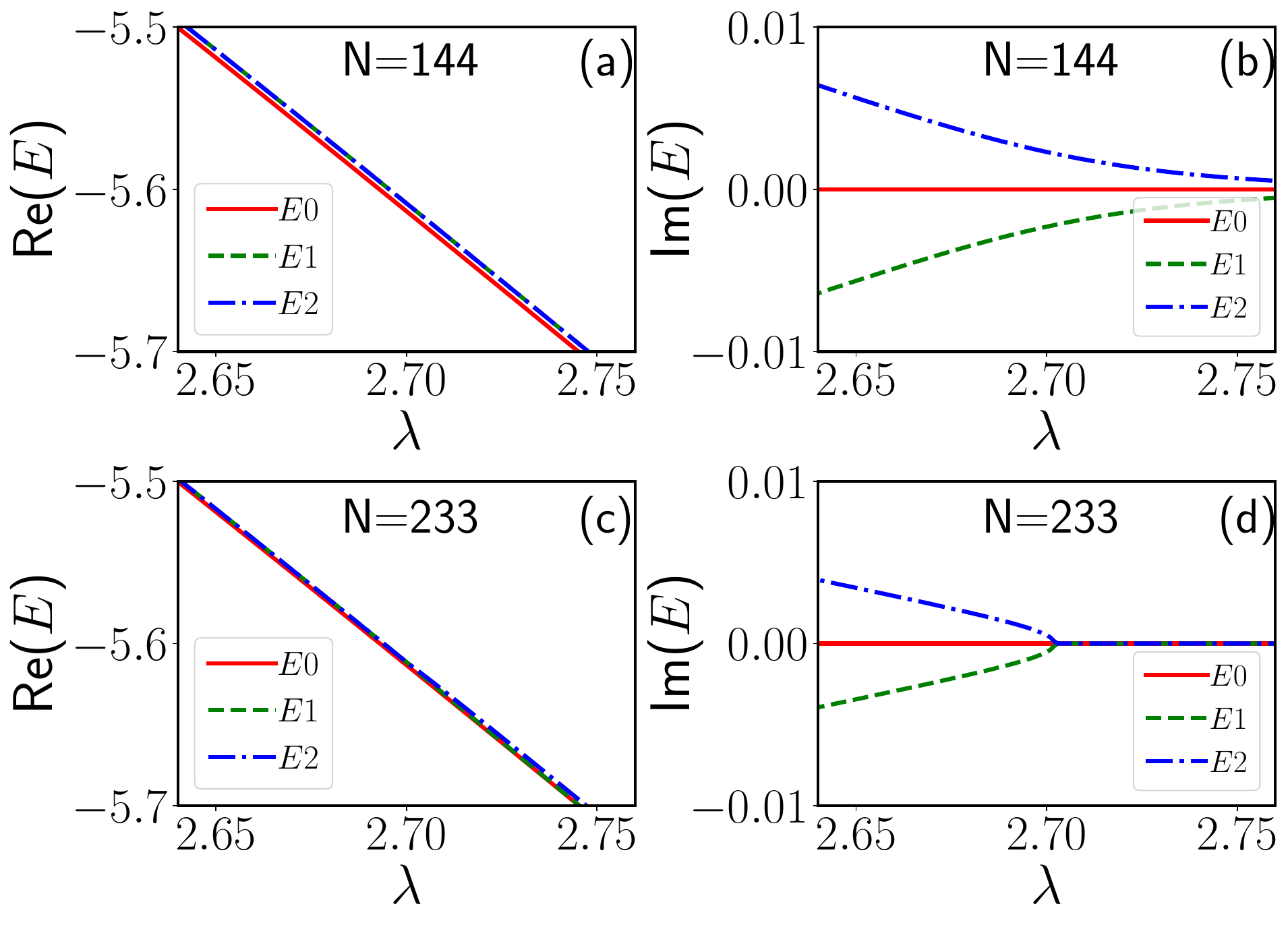}
	\caption{Spectrum of one-dimensional nonreciprocal AAH model for the three states with the lowest real part energies. 
	(a)(b) The real and complex parts of energies with respect to $\lambda$ for $N=144$ at $g=1$.
	(c)(d) The real and complex parts of energies with respect to $\lambda$ for $N=233$ at $g=1$.}
	\label{fig:spectrum}
\end{figure}

To explore the properties of the fidelity susceptibility around the $\mathcal{PT}$ transitions, we focus on studying the first-excited states.
For first-excited states, we observe that the fidelity susceptibilities display distinct scaling laws depending on whether the lattice comprises even or odd sites. 
For even lattices, the self-normal fidelity susceptibilities near the critical points persist in scaling as $N^{2}$ [c.f. Fig.\ref{fig:E1even}].
While, for odd lattices, we find that the real parts of the biorthogonal fidelities $F_{1}^{B}$ become $1/2$ at the critical point \cite{tu2023general} [c.f. Fig.\ref{fig:E1odd}(b) and Fig.\ref{fig:E1odd}(d)] independent on the lattice sizes,
indicating the biorthogonal fidelity susceptibilities $\chi_{F_{1}}^{B}$ would diverge as $\delta \lambda \rightarrow 0$ for arbitrary finite sizes. 
In contrast, the self-normal fidelities (which are always real) decrease [c.f. Fig.\ref{fig:E1odd}(a)]. It is observed that fidelity susceptibilities $\chi_{F_{1}}^{S}$ exhibit linear scaling with $\chi_{F_{1}}^{S} \propto N$ [c.f. Fig.\ref{fig:E1odd}(c)], with a slope of $0.9721 \pm 0.0050$.
Our findings thus indicate that the fidelity susceptibility serves as a reliable tool for detecting the phase transition of the AAH model beyond the real-energy regime. 
Furthermore, it suggests that it displays a distinct behavior in contrast to the ground-state fidelity susceptibility, indicating the presence of a novel scaling law for $\mathcal{PT}$ transitions.
The scaling laws are summarized in Table \ref{tab:scalingFS}. 
The distinctions can be understood from the energy spectrum shown in Figs.\ref{fig:spectrum}, where the ground state and the first-excited state with even sites, which undergo second-order phase transitions, change smoothly, while the first-excited state with odd sites undergoes a $\mathcal{PT}$ transition. 
For second-order phase transitions, the fidelity susceptibility in non-Hermitian systems is known to obey the scaling law given in Eq. (\ref{eq:FSscaling}) \cite{sun2022biorthogonal}. The unique scaling law we identified originates from these $\mathcal{PT}$ transitions.
Additionally, since the $\mathcal{PT}$ transition and the localization transition occur at the same point in this AAH model \cite{zhai2022nonequilibrium,jiang2019interplay, zhai2021cascade}, the ground-state fidelity can be used to describe the $\mathcal{PT}$ transition (specifically the critical point) for this model.

{\it Conclusion.-} In summary, our study focuses on exploring quantum criticalities and fidelity susceptibilities of both the ground state and first-excited states within the one-dimensional nonreciprocal AAH model. We show that the scaling laws of the fidelity susceptibilities in the AAH model can be divided into two subsequences depending on the odd/even properties of the denominators in the irrational number.
We demonstrate that the fidelity susceptibility serves as a reliable tool for characterizing phase transitions of the AAH model beyond the real-energy regime.

In the ground state of the AAH model, characterized by real eigenenergies, we observe that the scaling behavior of the fidelity susceptibilities is perfectly consistent with that of the Hermitian AAH model, exhibiting an equivalent correlation length critical exponent of $\nu=1$. 
In the first-excited state, the fidelity susceptibilities exhibit different scaling laws for the odd/even sites. 
For even lattices, the self-normalized fidelity susceptibilities exhibit a scaling behavior of $N^2$ akin to that observed in the ground state fidelity susceptibilities.
In contrast, on odd lattices, where the system has a $\mathcal{PT}$ transition, the self-normalized fidelity susceptibilities follow a linear scaling law, $\chi_{F_{1}}^{S}(\lambda_m) \propto N$, while the biorthogonal fidelity susceptibilities diverge.
The $1/2$ value of the biorthogonal fidelity has been observed in other systems \cite{tu2023general}. 
Exploring whether these scaling laws persist in the many-body AAH model and other systems would be an intriguing avenue for future investigation.

{\it Note added.-} After completing our work, we became aware of two related works focusing on characterizing the phase transition of a generalized AAH model using fidelity susceptibility \cite{ren2024identifying} and the scaling theory of the non-Hermitian disorder AAH model \cite{sun2024hybrid}, respectively.

{\it Acknowledgments.-} G.S. is appreciative of support from "the Fundamental Research Funds for the Central Universities, under the Grant No. NS2023055", the NSFC under the Grants No. 11704186, and the High Performance Computing Platform of Nanjing University of Aeronautics and Astronautics. C.Z., Z.C. and G.W. are appreciative of support from the Innovation Training Fund for College Students under Grant No. 202310287211Y.


\bibliographystyle{apsrev4-1}
\bibliography{NHAAH}


\end{document}